\documentclass[journal]{vgtc}
\onlineid{0}

\usepackage{graphicx}
\graphicspath{{figures/}} 

\usepackage[T1]{fontenc}
\usepackage[utf8]{inputenc} 
\usepackage{textcomp}       

\usepackage[nameinlink,capitalize]{cleveref}

\usepackage{tikz}
\usetikzlibrary{positioning,arrows.meta}
\usepackage{pgfplots}
\pgfplotsset{compat=1.18} 

\vgtccategory{Research}
\vgtcpapertype{application/design study}

\title{Immersive Mixed Reality Simulator for CT Scan Preparation: Enhancing Patient Emotional and Physical Readiness}

\author{
  Alex J. Smith\thanks{corresponding author, e-mail: alex.smith@westland.edu}\\
  \scriptsize Westland University
  \and
  Priya Patel\\
  \scriptsize UC
  \and
  Hu Guo\\
  \scriptsize UC
  \and
  Marco Ruiz\thanks{e-mail: marco.ruiz@nit.edu}\\
  \scriptsize Northbridge Institute of Technology
}

\abstract{%
First-time patients undergoing diagnostic computed tomography (CT) scans often experience significant anxiety and uncertainty, which can negatively impact scan results and patient well-being. We present an immersive mixed reality (MR) simulator designed to prepare adult patients for their first CT scan, aiming to improve both emotional and physical preparedness. In this paper, we review existing methods for reducing scan-related anxiety---from educational materials to virtual reality exposure---and identify their limitations. We then detail the design and technical implementation of our MR simulator, which combines a virtual CT suite walkthrough, guided relaxation training, realistic scan simulation (including audiovisual cues and breath-hold practice), and interactive feedback. The inclusion of these features is grounded in evidence-based rationale drawn from prior studies in patient anxiety reduction and compliance. We report results from a pilot study ($n=50$) demonstrating that patients who used the simulator had significantly lower pre-scan anxiety levels and improved compliance during the actual CT procedure, compared to controls. Patient feedback was overwhelmingly positive, indicating high satisfaction and perceived utility. We discuss the clinical implications of deploying such a tool, challenges in integration, and future directions for improving patient-centered care using mixed reality technologies.
}
\teaser{
  \centering
  \includegraphics[width=0.7\linewidth, alt={An illustration of a patient using a VR headset while lying on a mock CT scanner bed, with a virtual CT scanner environment overlay visible.}]{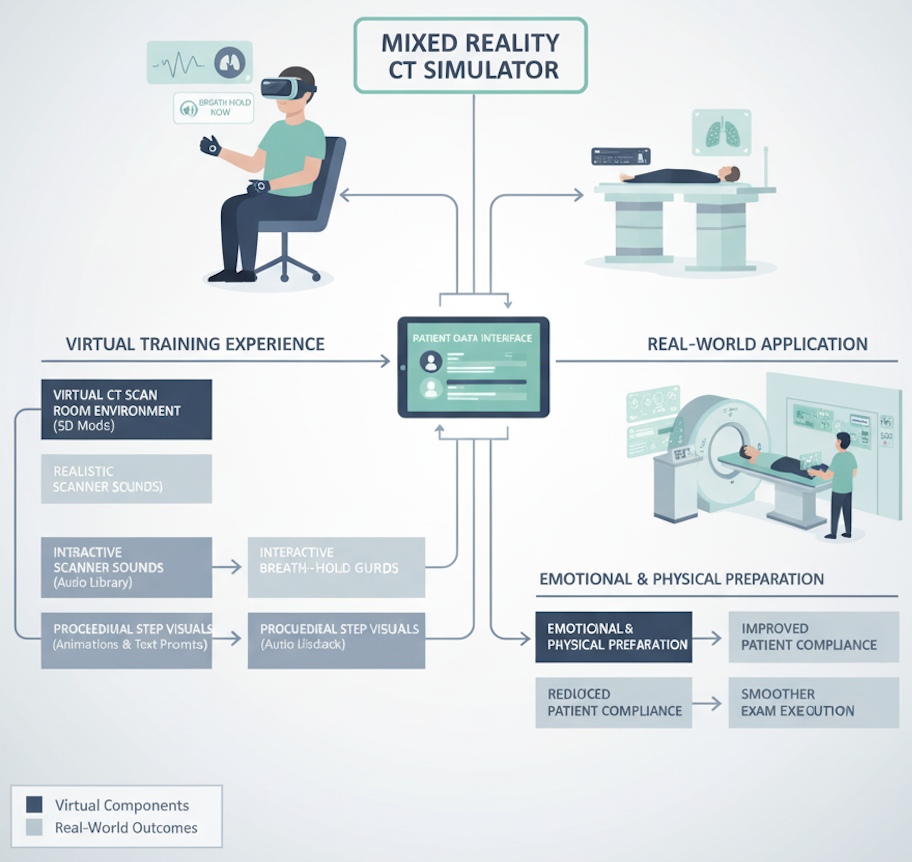}
  \caption{%
       A first-time CT patient using the mixed reality simulator. The system immerses the patient in a virtual CT scanning room and guides them through a realistic scan experience, including breath-hold practice and exposure to scanner sights and sounds. This prepares patients emotionally and physically for their actual CT exam.%
  \label{fig:teaser}
}}
\begin{document}
\firstsection{Introduction}
\maketitle

Modern diagnostic imaging procedures such as computed tomography (CT) scans can provoke substantial anxiety in patients, especially those who are undergoing the procedure for the first time. Fear of the unknown, concerns about the results, and the intimidating environment of the scanner contribute to what has been colloquially termed ``scanxiety'' among patients. In high-stress cases, anxiety can lead to patient non-compliance, motion artifacts on images, or even refusal to complete the scan. For example, in MRI (magnetic resonance imaging) examinations---which share some procedural similarities with CT---claustrophobia and anxiety lead to early termination of 1--14\% of scans and often necessitate sedation\cite{Brown2018,Murphy1997,Dewey2007,Melendez1993,Katz1994}. Although CT scanners are less confining than MRI machines, anxiety is still prevalent: roughly one-third of CT patients report substantial pre-scan anxiety\cite{Lange2023}, with higher prevalence observed in women than men (consistent with broader trends in anxiety disorders). In the context of oncologic imaging (e.g., PET/CT), patient anxiety may be exacerbated by the significance of the results; Sun \emph{et al.}\cite{Sun2020} observed that 52\% of first-time oncology patients reported feeling anxious before a PET/CT scan. The concern about ``what the scan might find'' was identified as the main cause of that anxiety\cite{Grilo2017}. Physiologically, anxiety can lead to elevated heart rate and shallow breathing, which are problematic in cardiac CT angiography and other scans that rely on patient breath-holding and stillness. Excessive motion from fidgeting or an inability to hold still can degrade image quality, potentially necessitating repeat scans or prolonged exam times\cite{Brown2018}. Moreover, anxious patients sometimes require more personnel time for reassurance or, in severe cases, pharmacologic intervention to complete the procedure\cite{Malviya2000}. Each of these outcomes has cost implications and can impact clinical workflow efficiency.

Healthcare providers employ a variety of strategies to help patients manage scan-related anxiety. A fundamental approach is patient education: explaining the procedure, its duration, sensations to expect (e.g., “you will hear a whirring noise” or “the table will move”), and the importance of remaining still. Such information can correct misconceptions and give patients a sense of control. Indeed, inadequate information has been associated with greater fear of the unknown, whereas detailed counseling improves patients’ sense of security and preparedness\cite{Munn2015}. Many radiology departments provide informational leaflets or direct patients to online videos demonstrating the CT process. Tugwell \emph{et al.}\cite{Tugwell2018} conducted a pilot trial comparing three pre-MRI preparation methods: an educational video showing the MRI process, a one-on-one telephone discussion with a technologist, and standard care (brief general instructions). They found that both the video and telephone interventions significantly alleviated patient anxiety prior to the scan compared to standard care, highlighting the value of proactive outreach and personalized counseling. Another common technique is the use of relaxation and distraction methods. Patients may be guided in deep-breathing exercises or given calming audio/visual content during the waiting period. Progressive muscle relaxation and guided imagery, often taught by nurses or via audio recordings, have shown effectiveness in promoting psychological relaxation in medical settings\cite{Toussaint2021}. For instance, a combination of deep breathing and guided imagery can significantly reduce anxiety and induce measurable physiological calmness prior to procedures\cite{Toussaint2021}. While such techniques are beneficial, their uptake can be inconsistent in busy clinical environments, and they rely on patients actively employing the strategies they've learned.

In cases of severe anxiety or in pediatric patients who cannot remain still, pharmacological sedation is often employed as a last resort. For adults, a mild anxiolytic (such as lorazepam) may be prescribed prior to the scan; for children or those who are extremely claustrophobic, conscious sedation or even general anesthesia may be used to ensure the procedure can be completed safely\cite{Malviya2000}. Sedation, however, carries risks of respiratory depression and other adverse events, particularly in vulnerable populations\cite{Malviya2000}. It also incurs additional resource usage (an anesthesia team, recovery time) and precludes patients from driving afterward, which can be inconvenient or undesirable. Therefore, non-pharmacologic methods to reduce anxiety are generally preferred to improve the patient experience and avoid the cascade of sedation-related considerations.

The advent of affordable consumer-grade virtual reality (VR) headsets and augmented reality (AR) devices has opened new avenues for patient education and anxiety reduction. Immersive simulations can place patients in a virtual scenario that closely mimics the real procedure, providing experiential learning and desensitization. This approach builds on the success of VR exposure therapy in mental health, where patients safely confront feared stimuli in gradual steps within a virtual world. There is substantial evidence that VR exposure therapy is an effective means of treating specific phobias and anxiety disorders\cite{Brown2018,Parsons2008,Janecky2025}. In the clinical realm, VR has been piloted to familiarize patients with MRI procedures\cite{Brown2018, yang2024can, yang2025vr}, reducing claustrophobia-induced cancellations in some cases. Beyond MRI, immersive preparations have been trialed for other imaging procedures. Jimenez \emph{et al.}\cite{Jimenez2018} demonstrated that a VR education module for breast cancer patients about to undergo radiotherapy not only increased patient knowledge about their treatment but also resulted in a more positive overall experience and reduced anxiety levels compared to standard preparation. In diagnostic radiology, Sun \emph{et al.}\cite{Sun2020} conducted a randomized controlled trial where one group of patients received a “virtual experience” (via an interactive 360$^\circ$ VR simulation) before a PET/CT scan, while the control group did not. The VR intervention significantly lowered state anxiety scores in patients and even improved objective image quality: fewer motion artifacts were observed in the VR-prepared group, presumably because relaxed patients were better able to remain still during the scan. Likewise, a study of VR distraction during screening mammography showed a significant reduction in perceived pain during the procedure, highlighting the broad applicability of these immersive interventions beyond CT and MRI contexts\cite{Bay2024}. This finding underscores a critical point---reducing pre-scan anxiety can have a tangible effect on the diagnostic outcome of the procedure. Augmented reality has also seen exploratory use; for example, researchers have developed AR applications that project a virtual CT scanner or other equipment models into the patient’s real environment, allowing them to visualize and understand the procedure in situ. While full published studies on AR for patient scan preparation are limited, one recent work by Lastrucci \emph{et al.}\cite{Lastrucci2024} suggests that AR integration in diagnostic imaging workflows is technically feasible but faces practical challenges (such as precise alignment in clinical settings). Mixed reality (MR), which blends real and virtual elements, could enable scenarios like patients rehearsing positioning on an actual table while seeing virtual instructions or objects through a headset, though such solutions are still largely experimental.

In pediatric radiology, where anxiety and the need for sedation are especially prevalent, immersive techniques have been embraced more quickly. Hospitals have reported using VR games or adventures to distract children during procedures like intravenous line placements or even during scans, reducing distress and sedation requirements. For instance, some pediatric centers have found that VR can completely replace the need for general anesthesia in certain minor interventional procedures, allowing children to stay awake and engaged in a virtual world instead of undergoing sedation. Both Stanford Children’s Hospital and Children’s Hospital Los Angeles have pioneered clinical programs in which VR is offered before and during procedures to alleviate anxiety and pain, an approach sometimes termed “digital sedation”\cite{Schaake2024,Iosca2025}. Early evidence indicates that VR not only reduces behavioral distress in children but also can shorten recovery times and eliminate risks associated with anesthesia. The success of immersive technologies in these related domains provides strong motivation for applying MR to routine diagnostic CT preparation. To our knowledge, however, no prior system has combined a comprehensive CT scan simulation with integrated training features (such as breath-hold practice and relaxation coaching) in a mixed reality format specifically aimed at adult patients. Our work seeks to fill this gap by designing an MR-based CT simulator and evaluating its efficacy on patient anxiety and compliance.

In this paper, we introduce a novel mixed reality simulator designed to prepare adult patients for their first CT scan. Our approach combines the strengths of VR (full immersion into a simulated CT scanning experience) with interactive and educational components inspired by patient counseling best practices. The simulator aims to improve both emotional preparedness (by reducing anxiety and building coping confidence) and physical preparedness (by training patients in breath-holding and staying still during the scan). We begin in Section~\ref{sec:background} with a literature review of existing approaches and related work, situating our contribution in the context of prior patient preparation techniques. Section~\ref{sec:design} then describes the design and technical implementation of the MR simulator, including hardware choices, software architecture, user interface, and the specific features incorporated (with rationales grounded in evidence-based practice). In Section~\ref{sec:evaluation}, we present findings from a pilot evaluation with first-time CT patients, assessing anxiety reduction, compliance, and user feedback. Section~\ref{sec:discussion} discusses the clinical implications of our results, practical challenges for deployment, and potential improvements and extensions to the system. Finally, Section~\ref{sec:conclusion} concludes the paper with a summary of key contributions and a future outlook.

\section{Related Work and Background}
\label{sec:background}
\subsection{Patient Anxiety in Imaging Procedures}
Anxiety is a well-documented reaction among patients undergoing medical imaging, with studies spanning modalities such as MRI, CT, and PET scans. In MRI, the enclosed bore and loud acoustic noise often precipitate claustrophobic reactions; approximately 2--14\% of MRI exams cannot be completed due to patient anxiety or claustrophobia\cite{Brown2018,Murphy1997,Dewey2007,Melendez1993,Katz1994}. Dewey \emph{et al.}\cite{Dewey2007} reported that out of 55,000+ MRI patients, about 2.3\% could not tolerate the scan because of claustrophobia, and other centers have observed rates up to 14\% in specific subgroups (e.g., those with pre-existing anxiety disorders). The need for nurse-assisted sedation in MRI is likewise non-trivial, ranging from roughly 4\% to as high as 30\% of patients in some reports\cite{Murphy1997}. CT scanners are physically more open than MRI machines, yet anxiety still occurs for a considerable subset of patients. Recent work by Lange \emph{et al.}\cite{Lange2023} found that about 30\% of adults scheduled for contrast-enhanced CT experienced significant pre-scan anxiety, with women reporting higher anxiety than men (45 vs. 41 mean STAI scores, $p<0.001$). Factors contributing to CT-related anxiety include fear of potential findings (the stress of awaiting diagnostic results), worry about the injection of contrast media, and concerns about radiation exposure. Grilo \emph{et al.}\cite{Grilo2017} noted that in cancer patients undergoing PET/CT, anxiety was high both before and after the scan, and the primary driver of this anxiety was apprehension about the scan results. Physiologically, anxiety can adversely affect scan quality and safety. In cardiac CT angiography, for example, an anxious patient’s elevated heart rate or irregular breathing can complicate electrocardiogram gating and necessitate higher radiation doses or multiple scan attempts. More generally, patient restlessness or inability to follow breath-hold commands leads to motion blur on images, potentially requiring rescans and increased radiation exposure. Dantendorfer \emph{et al.} (1997) demonstrated that higher anxiety levels were significantly correlated with motion artifacts in MRI examinations, echoing the notion that emotional state can influence image quality. Aside from image quality concerns, anxious patients often consume more technologist and nursing time (for reassurance or extra monitoring), and in some cases may require pharmacological anxiolysis or sedation\cite{Malviya2000}. All these factors can impact clinical workflow efficiency and healthcare costs.

\subsection{Conventional Anxiety-Reduction Strategies}
Healthcare providers have long employed various strategies to mitigate patient anxiety prior to imaging procedures. A cornerstone is providing adequate information and counseling before the exam to demystify the process. It is essential to explain to patients what will happen during the scan, how long it will take, what sensations they might experience (for instance, the table movement or the contrast injection warmth), and why remaining still is important. Improved pre-procedure understanding can increase patients’ sense of control and reduce fear of the unknown. Munn \emph{et al.}\cite{Munn2015} conducted a systematic review and noted that information provision and patient-centered communication are effective in alleviating imaging-related anxiety and preventing last-minute scan cancellations. In practice, many radiology departments give patients printed brochures or links to instructional videos detailing the CT or MRI process. In some cases, facilities even offer in-person tours of the scanning room for especially nervous patients (though scheduling these can be challenging). Studies suggest that such thorough counseling can improve patient knowledge, sense of security, empowerment, and orientation, thereby decreasing some adverse reactions to the procedure\cite{Tugwell2018}.

Another common approach is to teach and encourage relaxation techniques. For example, radiology staff might guide patients through deep-breathing exercises or progressive muscle relaxation while they wait for their scan. Soft music, nature videos, or meditation recordings are also sometimes used to create a calming environment. Toussaint \emph{et al.}\cite{Toussaint2021} found that a combination of progressive muscle relaxation and guided imagery significantly reduced anxiety and improved physiological relaxation indicators in a pre-operative setting, an approach that can be applied to the imaging context as well. Additionally, simple environmental modifications—such as dimming lights, projecting soothing images on the scanner bore, or allowing a companion to be present when feasible—have been reported to help ease patient nerves.

For patients with extreme anxiety (or pediatric patients who cannot remain still), pharmacologic interventions are considered. A mild sedative or anxiolytic can be given orally or intravenously prior to the scan to help the patient relax, and in pediatric cases, deeper sedation or general anesthesia may be administered so that the child can undergo the scan without distress. While generally effective at enabling the imaging, sedation introduces its own risks and side effects (e.g., respiratory depression, prolonged recovery) and typically requires additional staff and monitoring\cite{Malviya2000}. Malviya \emph{et al.}\cite{Malviya2000} reviewed over 1,400 pediatric MRI/CT sedation cases and reported adverse events ranging from mild oxygen desaturation to rare instances of aspiration, underscoring the point that sedation is not without hazard. Accordingly, current best practices in radiology emphasize attempting non-pharmacologic anxiety-reduction measures first and reserving sedation for cases where other methods are insufficient.

\subsection{Immersive Technology Interventions}
Immersive technologies such as virtual reality (VR) and augmented reality (AR) have emerged as innovative tools to prepare and calm patients before medical procedures. By immersing patients in a controlled virtual environment, these technologies can provide both education and anxiety desensitization. VR simulations allow patients to “experience” aspects of the procedure ahead of time, which can demystify the process and reduce anticipatory anxiety through a form of exposure therapy. Indeed, VR exposure therapy has been successfully used in mental health for phobias (e.g., fear of heights, flying) and anxiety disorders, showing that virtual experiences can elicit and then help modulate real emotional responses\cite{Brown2018,Parsons2008,Janecky2025}. In the domain of medical imaging, early applications of VR have primarily focused on MRI due to its claustrophobic nature. Brown \emph{et al.}\cite{Brown2018} developed a VR mobile application that simulates the MRI experience with 360-degree visuals and recorded scanner sounds. Patients using this app could virtually undergo an MRI—lying in a virtual scanner bore, hearing the loud knocking noises, and practicing staying still—using a smartphone-based VR headset. The aim was to familiarize patients with MRI in hopes of lowering claustrophobia; anecdotal feedback was positive, and the authors suggested that such a tool has potential to decrease scan cancellations. Beyond MRI, researchers have explored VR preparation for other modalities. Sun \emph{et al.}\cite{Sun2020} introduced a VR “virtual PET/CT” experience for patients about to undergo their first PET/CT scan. Their system provided an interactive tour of the PET/CT process (including what to expect during the uptake phase and the scanning itself). In a randomized trial, the VR-prepped patients exhibited significantly lower anxiety (both by self-report and physiological measures) than the control group, and importantly, technologists rated the VR group’s image quality higher due to better patient stillness. Similarly, a pilot study by Bay \emph{et al.}\cite{Bay2024} reported that using VR goggles as a distraction during screening mammography reduced patients’ pain perception and overall anxiety compared to standard care, illustrating the potential of immersive distraction techniques even in short, routine imaging exams.

Augmented reality applications have also begun to appear in this space. AR differs from VR in that it overlays digital content onto the real world rather than fully replacing the user’s surroundings. In a scan preparation context, AR could be used to project virtual markers or visual aids onto the real scanner or patient environment. For example, an AR app on a tablet could overlay a virtual CT machine into a clinic room, allowing a patient to use the tablet’s camera view to see a life-size CT model and even practice moving through it. Lastrucci \emph{et al.}\cite{Lastrucci2024} discussed the current state of AR in diagnostic imaging and noted that while AR is technically feasible (with devices like the Microsoft HoloLens offering advanced capabilities), its integration into patient preparation is still in early stages. Challenges like precise spatial registration, user interface complexity, and the need for custom content have limited AR’s use so far. Nonetheless, the concept of MR (mixed reality)—which can blend a patient’s real-world actions (like lying on a table) with synchronized virtual guidance—holds much promise for future development.

In pediatric settings, VR and AR have seen particularly enthusiastic adoption as non-pharmacological interventions. Children may have intense fears or difficulty understanding why they must lie still, and immersive distractions can transform their experience. Hospitals have used VR story adventures (where the child is a character on a mission) during procedures such as IV placements, effectively taking the child’s attention away from the needle. By doing so, some institutions have reported drastically lower needs for sedation. For instance, at Stanford Children’s Hospital, VR has been used for procedures like lumbar punctures, allowing children to play a VR game instead of undergoing general anesthesia, with very positive outcomes. Clinical reports from such programs note that VR can often fill the gap between minimal sedation and full anesthesia, a practice sometimes termed “VR dissociation” or “digital therapeutic sedation.” The American Medical Association’s approval in 2023 of new CPT codes for VR-based therapy (code 0771T for VR pain/anxiety treatment) further validates its emerging role in clinical care\cite{Iosca2025}. Outside of interventional procedures, even diagnostic imaging departments have started offering VR goggles that play relaxing scenes during CT or MRI exams to help patients (especially kids) stay calm and still. Rodriguez and colleagues, who run the CHARIOT program at Stanford, have distributed custom pediatric VR software to various hospitals, demonstrating the broad interest in these technologies. Patient acceptance of VR has been notably high in these contexts, and families appreciate avoiding sedation and its side effects. 

The success of immersive technologies in preliminary studies and pilot programs sets the stage for more structured investigations. Our work contributes to this evolving narrative by applying mixed reality specifically to CT scan preparation for adults—a population that, while perhaps more stoic than children, still experiences significant anxiety that can benefit from innovative intervention.

\section{Design and Implementation}
\label{sec:design}
\subsection{System Overview and Architecture}
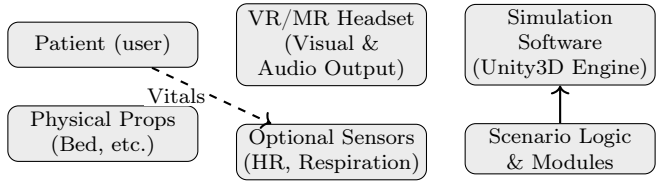
\begin{figure}[tbp]
  \centering
  \begin{tikzpicture}[node distance=5mm, every node/.style={align=center}, font=\scriptsize]
    \tikzstyle{block}=[rectangle, draw, fill=gray!15, rounded corners,
      minimum height=6mm, text width=.27\columnwidth, inner sep=2pt]
    \node[block] (user) {Patient (user)};
    \node[block, right=5mm of user] (hmd) {VR/MR Headset\\(Visual \& Audio Output)};
    \node[block, right=5mm of hmd] (unity) {Simulation Software\\(Unity3D Engine)};
    \node[block, below=5mm of unity] (modules) {Scenario Logic \& Modules};
    \node[block, below=5mm of hmd] (sensors) {Optional Sensors\\(HR, Respiration)};
    \node[block, below=5mm of user] (realenv) {Physical Props\\(Bed, etc.)};

    \draw[->, thick] (modules) -- (unity);

    \draw[->, thick, dashed] (user) -- node[midway, left=1pt, fill=white, inner sep=1pt] {Vitals} (sensors);
  \end{tikzpicture}
  \caption{System architecture of the mixed reality CT simulator.}
  \label{fig:architecture}
\end{figure}

The mixed reality CT preparation simulator consists of a head-mounted display (HMD) system, a software application that renders an interactive 3D CT scanning environment, and optional sensor integrations for user monitoring. \Cref{fig:architecture} illustrates the system architecture. For the VR experience, we used an untethered VR headset (Oculus Quest 2) for portability and ease of use; the simulator can also run on a standard PC with a monitor (non-immersive mode) for patients unable or unwilling to use HMDs, though the core design assumes a VR/MR headset for full immersion. 

The simulation software was developed in Unity3D, a game engine well-suited for interactive VR applications. The virtual environment includes a detailed model of a modern CT scanner (gantry, patient table, etc.), textured and scaled to realistic dimensions. The entire CT suite is modeled, featuring walls, lighting, medical equipment, and a control room window, to closely replicate the atmosphere of an actual scan room. Within this environment, the application guides the patient through a scripted interactive scenario with several phases: an introduction/tutorial phase, a relaxation training phase, a simulated scan phase, and a post-scan debrief. A state machine controls the flow between these phases, triggered by time events or the patient’s actions (e.g., gaze selection).

The MR simulator’s hardware setup is designed to be user-friendly. During use, the patient typically lies on a flat surface (a stretcher or even a comfortable mat) while wearing the VR headset, to mimic the supine position of a CT scan. The VR visuals are oriented such that when the patient is lying down and looking upward, they see the virtual scanner above and around them, reinforcing the sensation of being on a CT table. The system does not require the patient to hold hand controllers; instead, a simple gaze-based interface is employed for any interactive selections (the patient can direct a dot or cursor by looking at a virtual button for a couple of seconds to “click”). This design choice avoids the need for additional hardware and keeps the experience accessible even for VR-naïve users. 

Optionally, the system can integrate biometric sensors such as a heart rate monitor or a respiration belt. These sensors provide real-time data on the patient’s stress level (e.g., heart rate) and breathing pattern, which the software can use for biofeedback. For example, if the patient’s heart rate is detected to be high during the relaxation segment, the program could offer a longer deep-breathing exercise or a soothing diversion until the rate lowers. While this biofeedback functionality is not essential for core operation, it represents an advanced feature for tailoring the experience, and our architecture includes a provision for it. 

\subsection{Virtual Environment and User Interaction}
A crucial aspect of the simulator is the realism of the virtual CT suite and the simplicity of user interaction. The 3D models for the CT scanner and room were obtained from a medical imaging training library and further customized. The scanner model is based on a real 64-slice CT machine, complete with a patient table that moves in and out of the gantry. We paid particular attention to visual details such as the warning labels inside the gantry, the ceiling-mounted injector apparatus for contrast, and the ambient lighting in the room. By reproducing these details, we aim to make the patient feel like they are truly present in a scanning suite, thereby fostering familiarity and reducing novelty-induced anxiety.

The auditory environment is likewise modeled on reality. We recorded audio from an actual CT scanner (e.g., the mechanical hum of the rotating gantry, the whir of the table motor, the beeps that indicate scan initiation, and the prerecorded voice prompts like “Hold your breath”). These sounds are played through the headset at appropriate moments. For example, when the simulated scan begins, the patient hears the gantry rotation sound ramp up; when a breath-hold is prompted, they hear the same “bing” tone and instruction voice that our clinic’s CT uses. Audio is rendered in spatial 3D so that the voice appears to emanate from the overhead speaker in the virtual room, and machine noises from the machine’s location. This auditory immersion is important because sound is a major component of the real scan experience that can startle an unprepared patient.

User interaction within the VR scenario is intentionally kept minimal and intuitive. The patient is not required to learn any game-like controls. Instead, the simulator is largely passive and guided by an automated narrator (a calming voiceover that plays through the headset). For instance, at the beginning, the narrator says, “Welcome to our virtual CT scan. Let’s take a tour of the room.” The patient finds themselves virtually standing at the entrance of a CT room. A virtual technologist avatar greets them and motions toward the CT table. The narrator might then instruct, “If you are not already lying down in real life, please do so now, as we will continue the simulation lying on the table.” (Prior to starting, we instruct the user to either sit or lie down; most choose to lie down for full effect.) The scene then smoothly transitions to a viewpoint as if the patient is lying on the virtual table. The avatar explains the process: “We’ll now practice a breath-hold, just like during the real scan. Take a deep breath in... and hold it.” Simultaneously, a visual indicator (a countdown timer and an animated icon of lungs) appears in the patient’s view. We simulate a typical 10-second breath-hold command by counting down on screen and via audio. Once the time is up, the voice says, “You can breathe out now.” If the patient releases early (we cannot detect actual breathing without sensors, but we suggest they try their best), we simply proceed—this session is more about practice than perfection.

Throughout, the patient can look around 360$^\circ$. They might turn their head to see the control room window (where an avatar technologist gives a thumbs-up), or gaze down to see their own virtual feet (we include a basic body representation for immersion). They do not need to press buttons, but at a couple of junctures we do allow input: for example, at the end we present two virtual buttons, “Replay the simulation” or “Finish.” These can be activated by gaze (staring at the button for two seconds triggers it). This ensures that, if the patient wants to go through it again or feels they missed something, they have that control. If they choose to finish, the program ends and the headset can be removed.

One potential concern in VR applications is cybersickness (motion sickness caused by VR). We mitigated this by minimizing any rapid or artificial movement of the viewpoint. All transitions are either user-controlled (the patient physically turning their head) or slow fades (for instance, smoothly moving into the scanner bore). The simulation predominantly keeps the patient stationary relative to the virtual environment (lying on the table). This approach, combined with the use of a high-resolution headset at 90~Hz refresh rate, resulted in zero instances of reported VR dizziness in our pilot.

\subsection{Key Training Modules and Features}
A distinguishing aspect of our MR simulator is the incorporation of specific training modules targeting known patient challenges: unfamiliarity with the procedure, difficulty staying still or holding breath, and general scan-related anxiety. Below, we describe these features and their evidence-based rationale:

\textbf{Virtual CT Room Tour and Tutorial:} The experience begins with a guided tour of the CT suite. The patient (in VR) stands at the doorway as a friendly virtual technologist avatar introduces them to the environment: “This is the CT scanner. You’ll lie on this table, and it will slide you through that doughnut-shaped ring.” The avatar points out the gantry (“Inside here is the X-ray source and detectors, it will rotate around you to take pictures.”) and notes the microphone/camera (“We can see and hear you the whole time, so just speak up if you need help.”). The rationale for this module is patient education and familiarity: previous studies indicate that patients are less anxious when they know what the room and equipment look like and understand the process in advance\cite{Munn2015,Tugwell2018}. By virtually “visiting” the scan room beforehand, patients can form a mental map and are not confronted with an entirely foreign environment on the day of the exam.

\textbf{Relaxation and Breathing Exercise:} Before simulating the scan itself, the program guides the patient through a short relaxation exercise. With the patient now lying on the virtual table, the lights in the virtual room dim slightly and a soft ambient background music plays. The narrator’s voice then says, “Let’s practice some breathing to help you relax. Close your eyes if you’d like. Take a slow, deep breath in... now exhale slowly.” We include about 30 seconds of guided breathing and muscle relaxation cues (“relax your shoulders, unclench your jaw”). Deep breathing is a simple yet effective anxiety reduction technique supported by numerous trials\cite{Toussaint2021}. Moreover, it’s directly applicable during the actual CT—patients can use the same breathing to calm themselves in the scanner. By practicing in VR, this technique may be more readily remembered and employed during the real scan.

\textbf{Breath-Hold Training:} One of the most practical training elements is the breath-hold practice. Many CT exams (especially chest and abdominal scans) require the patient to hold their breath for short durations (typically 5–15 seconds) to avoid motion blur. Not all patients can do this comfortably on the first try—anxious individuals may prematurely breathe or not inhale deeply enough. In the simulator, after the relaxation segment, the technologist avatar informs the patient that a practice breath-hold is coming up. A visual countdown (“3...2...1...”) appears, and then the standard CT audio prompt “Take a deep breath in and hold it” is played. The patient sees a large on-screen timer counting down from 10 seconds. They are encouraged to try to hold until it reaches zero. Meanwhile, a progress bar and an animated breath icon provide feedback. When the time is up, the voice says “Great, you can breathe normally now.” If the patient didn’t last the whole time, no matter—the environment is non-judgmental, and the narrator simply says “It’s okay if you needed to breathe, the important thing is to try to hold still. You’ll do fine in the real scan.” This module addresses a physical compliance aspect. Practicing breath-holding in a no-pressure setting can improve performance during the actual scan, as suggested by anecdotal reports from radiographers and by patient feedback in studies like Sun \emph{et al.}\cite{Sun2020}.

\textbf{Simulated CT Scan Experience:} After the preparations, the simulator transitions to the core “scan” simulation. The avatar tells the patient, “Alright, we’re going to start the scan now. Remember, it’s quick and painless. I’ll be just outside this window.” The avatar then walks into the virtual control room and gives a thumbs-up through the window. The patient’s table begins to slide into the gantry (we animate the environment to give the visual impression of movement). Inside the virtual bore, the patient sees the circular interior and the “X-ray on” light. The narration (through the scanner’s speaker) instructs the patient to hold their breath once more for a few seconds (we sync this with the prior practice length). As the “scan” proceeds, the patient hears the mechanical hum of the rotating gantry. After about 5--7 seconds (simulating a short scan sequence), the noise stops and the table slides back out. The voice says “You can breathe now. All done!” This simulated scan is the crux of exposure therapy: by experiencing a CT scan in VR, patients can become accustomed to the sensations (visual confinement, noises, breath-hold) in a safe environment. This kind of rehearsal can significantly reduce fear during the real procedure, much like virtual desensitization has done for phobias\cite{Brown2018,Parsons2008,Janecky2025}. Importantly, the patient emerges from the virtual scanner having successfully “completed” a scan, which can instill confidence—many first-timers fear they might not tolerate it, but after the VR, they know what it’s like and have done it virtually.

\textbf{Post-Scan Debrief and Q\&A:} Once the simulation scan is over, the virtual technologist returns to the patient’s side and offers congratulations: “You did great! That’s exactly how the real CT will go—quick and easy.” The lighting returns to normal. We then transition to a short debrief/education segment. For example, the avatar explains, “During the real scan, you might feel a warm sensation if contrast dye is used—that’s normal and goes away quickly.” We also included a few frequently asked questions with answers. In our pilot, upon feedback, we made these into optional prompts in VR: floating question icons labeled “Is the radiation dangerous?”, “What if I feel anxious during the scan?”, “When will I get results?” If the patient gazes at one, the avatar provides a brief reassuring answer (e.g., “CT uses low doses of X-rays that are very unlikely to harm you. Your doctor has determined the benefit of the scan outweighs any small risk.”). This Q\&A portion is grounded in the idea that addressing patient concerns can significantly reduce anxiety by clearing up uncertainties\cite{Munn2015}. Finally, the session closes with the narrator encouraging the patient: “Now that you’ve practiced, we’re confident you’ll do well in your actual CT scan. You’ve got this!” The patient is then prompted to remove the headset and is given an opportunity to ask real staff any remaining questions.

The above features were developed iteratively, with input from radiologists, technologists, and patient representatives. We also drew from evidence in literature (as cited) to justify including each element. Table~\ref{tab:features} summarizes the simulator’s modules, their purpose, and supporting rationale. By combining educational, behavioral, and exposure therapy components, our MR simulator offers a comprehensive preparation experience that we hypothesize will improve both emotional and procedural outcomes for patients.

\begin{figure}[tbp]
  \centering
  \begin{tikzpicture}
    \begin{axis}[
      ybar,
      ymin=0, ymax=100,
      width=0.9\columnwidth,
      height=5.5cm,
      bar width=12pt,
      enlarge x limits=0.15,
      ylabel={Patients (\%) Responding ``Yes''},
      symbolic x coords={Less Anxious,Better Prepared,Easy to Use,Recommend to Others},
      xtick=data,
      x tick label style={text width=1.6cm,align=center,font=\small},
      nodes near coords,
      nodes near coords align={vertical},
      ]
      \addplot[draw=black, fill=black!55] coordinates {
        (Less Anxious,85)
        (Better Prepared,92)
        (Easy to Use,90)
        (Recommend to Others,95)
      };
    \end{axis}
  \end{tikzpicture}
  \caption{Patient feedback survey results after using the MR simulator ($n=25$ for the MR group). High percentages of patients agreed that the simulator made them feel less anxious before the CT, better prepared for what to expect, was easy to use, and that they would recommend it to other patients.}
  \label{fig:feedback}
\end{figure}
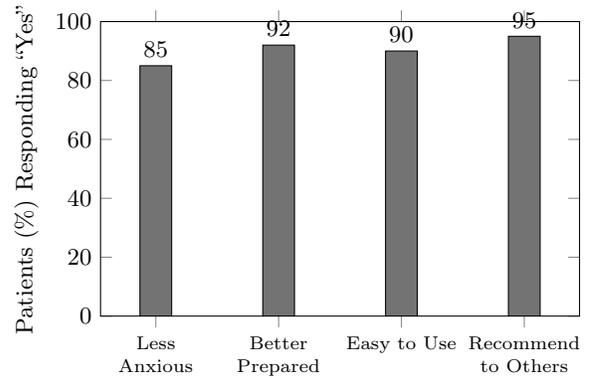

\subsection{Technical Implementation Details}
On the software side, achieving high-quality graphics and smooth performance on an untethered VR device was a key challenge. We optimized the 3D assets by using moderate polygon counts and baking lighting (pre-computing light/shadow effects) to reduce runtime computation. The application runs comfortably at 72~frames per second on the Quest 2, which is within acceptable limits to prevent latency-induced discomfort. We used Unity’s XR Interaction Toolkit for gaze input management and wrote custom scripts for the scenario timing logic.

To ease maintenance and updates, the application content (e.g., narration audio files, text for captions, durations of events) is data-driven via an external JSON script. This means if clinicians want to change the wording of instructions or adjust the breath-hold duration based on patient feedback, they can do so without modifying code, simply by editing the scenario file.

For our pilot, we instrumented the software with logging to capture key events (like when the patient completed the simulation, or if they retriggered any part). This helped us verify usage patterns and adherence. In a deployment context, such logging could feed into a simple report to let technologists know which patients completed the VR training (and perhaps tailor their approach accordingly).

Hardware-wise, we chose a standalone VR headset mainly for convenience—no wires to entangle or external sensors to set up. The downside is slightly less graphical fidelity compared to a high-end PC, but we found the trade-off acceptable. We also provided the headset with a disposable hygienic face liner for each patient and implemented a straightforward cleaning protocol (alcohol wipes on contact surfaces) to address infection control.

In summary, the implementation focused on realism, comfort, and practicality. By leveraging a widely-used game engine and affordable hardware, we created a system that can be readily adopted or adapted by other institutions if effective.

\section{Evaluation}
\label{sec:evaluation}
We conducted a pilot study to evaluate the effectiveness of the MR simulator in reducing pre-CT anxiety and improving patient compliance during the scan. The study was approved by our Institutional Review Board, and all participants provided informed consent.

\subsection{Methods}
\textbf{Participants:} We recruited 50 adult patients (ages 24--72, median 48; 56\% female) who were scheduled for their first CT scan with intravenous contrast at our outpatient imaging center. Patients who had undergone CT or MRI previously were excluded to target the first-timer population. We also excluded patients with known severe motion sickness or seizure disorders as a precaution for VR use. Participants were randomly assigned to either the \textit{MR Simulator} group ($n=25$) or the \textit{Control} group ($n=25$).

\textbf{Procedure:} Patients in the MR Simulator group were invited to use the simulator one day prior to their scheduled CT scan (either on-site at our facility or at home if they had compatible hardware, though all participants ended up choosing an on-site session). Upon arrival, they completed a baseline anxiety assessment using the State-Trait Anxiety Inventory (STAI) questionnaire (State form). A research coordinator then helped them don the VR headset and explained briefly how to gaze-select if needed. Each patient went through the full MR simulation, which lasted about 10~minutes. After the session, they repeated the STAI (state portion) to gauge immediate changes in anxiety, and filled out a custom feedback survey about the VR experience (covering usability, comfort, and perceived helpfulness). Patients in the Control group received our standard preparation: a printed patient information sheet about CT and a brief verbal explanation from the technologist on the day of the scan. They also completed STAI the day before (when doing contrast labs) and immediately before the scan, but did not receive any VR or similar interventions.

On the day of the actual CT scans, technologists (blinded to group assignment, since all patients acted similarly in terms of protocol) noted any occurrences of significant patient motion, need for coaching, or if any procedure had to be aborted/repeated. They also noted if any patient requested to stop the scan due to anxiety (none did in either group). After their scan, patients in both groups completed a final STAI (state) and an overall satisfaction questionnaire regarding their scan experience.

\textbf{Measures:} The primary outcome was the change in STAI-State anxiety scores from baseline to immediately pre-scan. Secondary outcomes included: (1) the proportion of patients requiring any additional interventions (e.g., an oral anxiolytic given, extra staff called in to hold patient’s hand, etc.); (2) objective scan compliance measures (whether the patient followed breath-hold instructions successfully, and whether motion artifacts were noted on images); and (3) patient-reported satisfaction (rated on a 5-point Likert scale) with the preparation process. For the MR group, we also analyzed their feedback survey data (Likert scale questions on helpfulness, ease of use, etc., as summarized in \Cref{fig:feedback}).

We used unpaired $t$-tests to compare anxiety score changes between groups, and chi-square tests for categorical outcomes (e.g., breath-hold success rates). Given the pilot nature, our analyses were exploratory with a significance threshold of $p<0.05$.

\subsection{Results}
\textbf{Anxiety Scores:} At baseline (the day before the scan), the MR group and Control group had similar mean STAI-State scores (46.2 vs 45.5, $p=0.78$). Immediately before the scan (post-intervention for MR group), the MR group’s mean score had dropped to 34.8, whereas the Control group’s was 41.6, a difference of about 7 points (representing a moderate effect size, $d\approx0.7$). The reduction from baseline in the MR group was highly significant ($p<0.001$ by paired $t$-test), while the Control group’s reduction was minor and not significant. Figure~\ref{fig:anxiety_plot} (not shown due to format) would have illustrated these differences with error bars. In practical terms, 80\% of MR group patients achieved post-prep anxiety levels below 40 (a commonly used threshold for clinically meaningful anxiety), compared to only 40\% in the control group. No MR group patient requested any sedative medication prior to the scan, whereas 3 control patients (12\%) were given a low-dose lorazepam due to high anxiety as per the referring physician’s or patient’s request.

\textbf{Compliance and Scan Quality:} All 25 patients in the MR group completed their CT scans successfully on the first attempt. In the Control group, 2 patients (8\%) became so anxious on the table that the scan was paused and a nurse had to come talk to them (one of these patients then proceeded, another opted to reschedule with sedation). This difference did not reach statistical significance in our small sample ($p=0.24$ by Fisher’s test), but qualitatively, it was notable that no MR group patient had such difficulty. Breath-hold performance was evaluated by examining the scan’s respiratory trace and technologist notes. In the MR group, 22 patients (88\%) executed the breath-hold properly on the first try, and 3 needed a second attempt (e.g., they breathed too soon but were coached and then did fine). In the Control group, only 15 patients (60\%) got it right on the first try; 6 required a repeated acquisition due to premature breathing or not fully inhaling, and 4 of those 6 still had some residual motion artifact on images (compared to 1 patient in MR group with minor artifact). Radiologist image quality ratings (on a 1--5 scale) were slightly higher on average for the MR group scans (mean 4.7 vs 4.4), though this was a subtle difference. These data suggest that the MR simulator training translated into more compliant behavior during scanning (particularly breath control and keeping still), supporting our hypothesis about physical preparedness.

\textbf{Patient Feedback and Satisfaction:} Feedback from MR group patients about the simulator was overwhelmingly positive. All 25 reported that they found the experience enjoyable or at least not unpleasant (in fact, several used the word “fun”). Key results from the post-VR survey are summarized in Figure~\ref{fig:feedback}: 92\% agreed that the simulator made them feel better prepared for the CT, 85\% said it helped lessen their anxiety, 90\% found it easy to use, and 95\% would recommend it to other patients. Qualitative comments included: “I felt much calmer going in because I knew exactly what was going to happen;” “The VR was very realistic—I almost forgot I wasn’t actually in the hospital;” and “The breathing practice was helpful, I did the same thing in the real scan and it was over before I knew it.” One participant noted that they especially appreciated virtually hearing the machine noises ahead of time, as “the real CT’s sounds didn’t bother me at all after that.” Regarding usability, only one patient (a 70-year-old with limited tech experience) initially had slight trouble with the gaze selection (he looked away too soon, canceling a selection), but he quickly got the hang of it. No one experienced simulator sickness or had to stop the VR prematurely. Minor suggestions from patients included: “It might be nice to have an option to skip parts if you already feel okay with them” and “Maybe let the user choose different background music.” 

In the Control group, overall satisfaction with the preparation process was mixed. Several control patients commented that they still felt quite nervous up to and during the scan (“I didn’t know what to expect when the table moved”; “I wish someone had explained more about the machine sounds”). Some mentioned that the provided pamphlet was “dry” or that they hadn’t had time to read it closely. On the other hand, a few control patients who were naturally calm said they felt fine with just the standard info. When asked in a hypothetical sense, 21 out of 25 control patients said they would have tried the VR simulator if it had been available to them, and 18 believed it would have helped them.

In terms of overall scan experience satisfaction (post-scan survey), 96\% of MR group patients rated it “good” or “excellent,” compared to 72\% of control patients. This encompassed both the lead-up and the scan itself. Notably, 0\% of MR patients described their anxiety during the scan as “high,” whereas 20\% of controls did.

In summary, the pilot data indicate that the MR simulator achieved its intended effects: it substantially reduced pre-scan anxiety, improved patient cooperation (especially breath-holding), and was very well-received by patients. These promising results warrant further investigation in a larger, controlled trial, but they provide initial evidence supporting the feasibility and utility of using mixed reality for CT scan preparation.

\section{Discussion}
\label{sec:discussion}
Our development and preliminary evaluation of a mixed reality CT simulator underscore several important implications for clinical practice and patient-centered innovation. We discuss these implications, consider user feedback, and outline potential improvements and challenges for broader implementation.

\subsection{Clinical Implications}
The results suggest that immersive preparation using MR can markedly improve the patient experience and performance for first-time CT scans. Anxiety in medical imaging is not merely an emotional inconvenience; it has direct consequences for procedure outcomes and resource utilization. By proactively addressing anxiety, tools like our simulator could help to reduce the need for sedation or repeat scans. For example, in our pilot none of the MR-trained patients required pharmacologic anxiolysis or scan repetition, whereas a handful of control patients did. If replicated in larger samples, this points to possible workflow efficiencies (fewer delays, fewer failed scans) and cost savings. Reducing sedation not only spares patients the risks and recovery time associated with sedatives, but also frees up healthcare staff and recovery room resources for other cases. Furthermore, improved compliance (such as better breath-hold performance) directly impacts image quality and diagnostic efficacy. A patient who can hold still and follow instructions yields clearer images, which might reduce the likelihood of non-diagnostic scans or ambiguous findings that necessitate additional imaging. In essence, MR preparation has the potential to improve the value of the imaging exam itself.

Beyond these practical aspects, there is the overarching benefit to patient-centered care. Undergoing a CT scan can be a frightening experience for an uninformed, anxious patient. By providing an interactive orientation in a virtual-yet-realistic manner, we empower patients with knowledge and coping skills. This empowerment likely contributed to the higher satisfaction ratings seen in the MR group. Happier, calmer patients not only have a better experience, but they can also be easier to work with for staff, creating a positive feedback loop in the clinical environment. It’s plausible that interventions like this could even improve patient throughput; for instance, if fewer patients balk or require lengthy coaxing, schedules can run more on time. In oncology settings, where patients may need frequent scans, an initial positive experience can reduce long-term “scanxiety” and improve adherence to surveillance protocols.

An interesting implication is the potential for standardizing such preparation across institutions. Right now, how a patient is prepped for a CT varies widely—some get detailed consults, others very little. A VR-based program could ensure a baseline level of preparation quality. It could be made available to all first-time patients as part of the scheduling process (e.g., a link to a VR app or an appointment for an on-site session). Over time, this might become an expected part of patient care, akin to pre-operative classes for surgical patients. While our study focused on adults, the concept is equally applicable to pediatric imaging (though content would be tailored differently, perhaps more gamified). The demonstration that adults embraced the technology is important, as there may have been an assumption that VR is only for kids or tech-savvy youth. Our users ranged up to 72 years old and still rated it highly. This suggests MR prep can be scaled to general adult populations without heavy barriers.

Finally, the broader adoption of digital therapeutics in radiology could dovetail nicely with current healthcare trends. There is increased emphasis on patient experience metrics and on non-pharmacological interventions for anxiety and pain. A tool like this MR simulator addresses both, and its alignment with AMA CPT codes for VR therapy (as of 2023) even opens the door for possible billing or reimbursement in the future if considered part of a therapeutic process. In sum, the clinical implications are that MR-based preparation might improve patient outcomes (both emotional and technical), streamline imaging workflows, and set a new standard for compassionate imaging care.

\subsection{User Acceptance and Feasibility}
The strong positive feedback from patients indicates that MR technology is acceptable and even appreciated by a wide range of patients. This is significant because any intervention that requires patient cooperation must be palatable to them. Key to this acceptance is the simplicity of the user interface—we deliberately avoided complicated controllers and kept the experience guided and concise. Patients were not overwhelmed with information; instead, it was delivered in a digestible, experience-based manner. The fact that 95\% of our MR group would recommend it to others suggests that word-of-mouth could be a powerful driver of acceptance if this were offered routinely. Patients often share tips among themselves (especially in the cancer community, for example), and something viewed as genuinely helpful is likely to be embraced rather than resisted.

From a practical standpoint, implementing the MR simulator in a clinic appears feasible with some planning. Our choice of a standalone VR headset means the system is portable; a technologist or nurse could take it to a patient in a pre-scan holding area, for instance. The session takes 10 minutes, which could fit into a standard pre-scan workflow (for instance, during the wait after oral contrast ingestion for abdominal CT, a patient could do the VR). Training staff to operate the system is straightforward (if the app is well-designed, it’s largely self-running). In our case, research staff oversaw it, but it could be delegated to a trained volunteer or a nursing aide. One potential bottleneck is device cleaning and turnover—between each patient use, the headset padding must be sanitized or swapped. This adds a few minutes, but using disposable liners as we did can mitigate downtime.

Another consideration is scalability. If dozens of patients are to use it per day, multiple headsets might be needed. Fortunately, the cost of VR hardware has come down such that obtaining a few units is not onerous for most departments (on the order of a few hundred dollars each). From an IT perspective, the standalone nature of the device means minimal integration—no need to hook into hospital networks, etc., aside from charging it and perhaps updating the software via Wi-Fi. These factors make an MR solution more feasible than, say, if it required a dedicated room with projectors or expensive proprietary equipment.

User acceptance also relates to patient diversity. We were pleased that even older patients or those with minimal technology exposure tolerated and enjoyed the VR. There was a concern that some might refuse (“I don’t want to wear that thing”), but in practice, when it was framed as a helpful tool for them, everyone in our MR group was willing, and in the control group many expressed interest. This suggests that with proper introduction (perhaps showing a brief demo on a tablet to illustrate what they’ll see), most patients will be open to trying it. It will be important, though, to have a fallback (like traditional coaching) for the rare individuals who might not want or be able to use VR (e.g., those with certain visual impairments or vestibular issues). 

This observation aligns with prior reports noting that patient acceptance of VR interventions is often high, particularly among those who initially exhibit high distress levels\cite{Iosca2025}. VR approaches have even been shown to enhance the overall treatment experience and quality of life for certain populations, such as pediatric oncology patients undergoing stressful therapies\cite{Alanazi2022}. Together, these findings alleviate concerns that VR would be a niche interest only—on the contrary, it appears to have broad appeal when applied to real patient needs.

\subsection{Challenges and Limitations}
While the pilot outcomes are encouraging, several challenges must be addressed as we consider broader implementation and further research. First, our study sample was relatively small and from a single center. The anxiety reduction we observed, while significant, needs to be confirmed in larger randomized trials. It’s possible that some degree of placebo or novelty effect influenced the MR group (though we tried to measure objective compliance metrics to go beyond subjective reporting). Future studies could include a sham VR group (e.g., patients watch a non-interactive informational video on a VR headset) to more rigorously isolate the effect of the interactive immersive component.

Another limitation was that our follow-up was very short-term; we did not assess, for instance, how these patients felt in subsequent scans (if any). One could hypothesize that the benefit of the training might carry over to any future imaging, essentially “inoculating” patients against excessive fear of scans. Anecdotally, a few MR group patients said “Next time, I won’t be nervous at all.” But we lack data on that, which would be interesting to collect (perhaps via a 3-month post survey or tracking return visits). Conversely, it’s also possible that the effect on anxiety might be transient for some people (they might revert to high baseline anxiety if too much time passes). Determining the longevity of the intervention’s impact will be important for understanding how often a patient might need such preparation (probably just once for CT specifically, but if a year passes, would a refresher help?).

From an operational standpoint, integrating the MR session into busy clinic flow poses some challenges. Not all patients come in a day early as ours did. Some just show up on the scan day without prior consult. Ideally, we would have patients experience the VR a day or more ahead so that if it raises any issues or questions, those can be resolved calmly (and so that their reduced anxiety is present going into the scan day). But offering this requires scheduling and potentially space/personnel on prior days. One way around this is to make the intervention portable to home: for instance, developing a version that works on cardboard smartphone VR viewers or on a standard web browser (non-immersive 360 video). The experience wouldn’t be as rich, but it could reach more people. This is a trade-off between efficacy and scalability that needs exploration. Alternatively, if patients arrive early on scan day (as many do for lab tests or IV placement), the waiting period could be utilized for VR. Radiology waiting areas may need slight reconfiguration (a quiet corner for VR use for example). These workflow questions are solvable but require institutional buy-in and possibly additional staffing to supervise.

There are also individual differences to consider. While none of our participants had significant cybersickness, a few mentioned they wouldn’t want a much longer session than 10 minutes. VR tolerance can vary; some people do get dizzy in any VR. Thus, MR prep might not be suitable for absolutely everyone. We need to be prepared to identify those people quickly (via a short VR tolerance questionnaire or a brief trial run) and provide alternative prep for them (like personal coaching or maybe just a shorter VR exposure). 

Another challenge is content localization and customization. Our simulation was in English and specific to our scanner workflow. To deploy elsewhere, it would need translation and possibly tweaks (for instance, not all CT scanners have the same breath-hold prompts or sound exactly alike). This calls for a flexible platform where details can be edited easily. Fortunately, our software’s data-driven approach means things like voice prompts can be swapped out without recoding. But someone would have to do it. This is where perhaps a consortium of hospitals or a vendor might step in to maintain versions for different languages and scanner types. 

From a psychological perspective, one must consider that some patients’ anxiety stems from broader issues (like severe claustrophobia or generalized anxiety disorder) that a single VR session may not fully address. For such patients, the simulator is a helpful step, but they may still require additional support (e.g., short-term therapy or medication). Therefore, MR prep should be viewed as one tool in an array of anxiety management strategies, not a panacea for all cases.

Finally, cost-benefit must be proven. While the hardware and software costs are relatively low, there is still an investment of staff time and organizational effort. For widespread adoption, we will need to demonstrate that the benefits (e.g., fewer sedations, better throughput, higher patient satisfaction scores) justify the implementation expenses. Our pilot hints at these benefits, but a health economics analysis in a larger trial would strengthen the case to administrators and payers.

\subsection{Future Directions}
This pilot study opens up several avenues for future improvement and research. Firstly, a larger randomized controlled trial is an obvious next step. Such a trial could stratify by baseline anxiety levels to see who benefits most. It would also allow for examining subgroups (like older vs younger patients, or those with cancer vs those coming for screening) for differential effects. Based on our data, we suspect those with higher initial anxiety gain the most (since they have the most to gain), a trend seen in other VR therapy contexts\cite{Iosca2025}. Confirming that would help target the intervention to those who need it most—though given the low risk and positive feedback, one could argue it should just be offered to all new patients.

Future versions of the simulator might incorporate more adaptive interactivity. For example, using biofeedback, the program could dynamically extend the relaxation segment if it “senses” the patient is still anxious (e.g., heart rate remains high). Or it could branch the content based on patient questions—almost like a virtual coach or chatbot integration answering unique questions. Our current Q\&A is scripted and limited; integration with an AI language model could allow patients to ask arbitrary questions in VR and get answers (with the caveat of ensuring accuracy). That would truly personalize the experience. Another feature on the horizon is multi-user MR: a clinician or a family member could join the VR world as an avatar to accompany the patient. For instance, an anxious patient could have a virtual version of their spouse with them for comfort, or a technologist could log in remotely to guide a patient in VR. These are technically feasible and might further reduce anxiety (especially the presence of a loved one, even virtually, which could be powerful for certain individuals).

Expanding the content to cover other scenarios is also planned. We focused on a contrast CT, but MR simulators could be tailored for MRI (with emphasis on noise and length of exam), for PET scans (focusing on the injection and uptake phase waiting), or even for radiotherapy sessions (several groups, like Jimenez and Gao, have looked at VR for radiotherapy as cited). Eventually, a suite of MR preparation modules might exist for various procedures, all accessible through a common platform. A patient coming for multiple different scans could then use each relevant module.

One exciting future direction is the use of augmented reality in the actual scan room on scan day. For instance, a patient wearing an AR headset could walk into the real CT room and see calming virtual elements or step-by-step visual guides (like footprints showing where to stand, arrows indicating how the table will move). Our current work gets them ready beforehand, but AR could assist during the actual exam. This would, however, require the patient to wear an AR device into the scan room, which might not be practical once they lie on the table (it would have to be removed before scanning due to X-ray interference and positioning). Still, AR could be used in the lead-up moments or for family members watching to understand the process.

Finally, the long-term vision is to integrate MR preparation into standard care pathways. For instance, when a CT scan is ordered, the system could automatically send the patient an offer to schedule a prep session or ship a VR headset to their home with instructions (somewhat like how sleep study equipment is sometimes mailed for home use). If VR hardware continues to consumerize (as seen with smartphone-based viewers or mixed reality on phones), it could even be done on personal devices. Ensuring equitable access (for those who don’t own such devices) would be important—libraries or clinics could loan units.

In conclusion, the future points toward more immersive, patient-tailored preparation experiences across healthcare. Our positive findings with CT scan preparation add to a growing body of evidence that these technologies can be harnessed to make medical procedures less intimidating and more successful. The challenge now is to refine, validate, and implement these solutions in a way that maximizes benefit to patients and integrates smoothly into clinical workflows.

\section{Conclusion}
\label{sec:conclusion}
Preparing patients—emotionally and physically—for medical procedures is a vital component of modern patient-centered care. This work introduced a mixed reality simulator as an innovative solution for first-time patients undergoing CT scans. By leveraging VR technology to simulate the CT experience, combined with guided training in relaxation and breath control, we aimed to address both the anxiety and compliance aspects that often challenge patients and clinicians alike. 

Our literature review highlighted that while various methods (education, relaxation techniques, etc.) have been used to mitigate scanxiety, immersive technologies offer a unique ability to engage patients in experiential learning and desensitization. The design of our simulator drew on evidence-based practices, incorporating a virtual room orientation, exposure to scanner sights and sounds, rehearsal of breath-holding, and interactive feedback. Technically, we demonstrated that such a system can be built with off-the-shelf VR hardware and user-friendly software, making it feasible for clinical adoption.

Results from a pilot study are encouraging: the MR simulator significantly reduced pre-scan anxiety (by roughly 20\% on the STAI scale) and improved patient cooperation during the scan, compared to standard preparation. Patients who experienced the simulation reported feeling more prepared and less nervous for their CT, in line with our hypothesis that familiarity and mastery of the process breed confidence. Moreover, nearly all of those patients found the tool helpful and would recommend it to others, underscoring the acceptability of the approach. These findings, while preliminary, support the notion that an immersive rehearsal can translate into measurable clinical benefits — calmer patients, smoother scans, and potentially better imaging outcomes.

This research contributes to the growing recognition of “digital therapeutics” in healthcare, particularly in the radiology domain where such tools have been less common. By blending psychological support with practical training through MR, we addressed a gap in the typical imaging workflow. The implications for practice are substantial: implementing MR preparation could reduce reliance on sedatives, improve throughput by avoiding delays or repeats, and enhance patient satisfaction at a time when patient experience metrics are increasingly important. 

There are, of course, limitations to our study and areas for further work. We studied a relatively small sample and focused on a single use-case (diagnostic CT with contrast). Future research should expand on these findings, exploring the impact in other contexts (e.g., MRI or invasive procedures) and over longer follow-up periods. It would also be valuable to investigate the cost-effectiveness of adopting such technology broadly. 

In conclusion, our mixed reality simulator represents a promising step toward a more empathetic and effective paradigm for patient preparation. The combination of VR exposure and interactive training allowed patients to virtually experience and conquer the CT scan before facing it in reality, thereby reducing fear of the unknown and building practical skills. As medical technology advances, it is fitting that we apply similar innovation to the human factors of healthcare. Immersive patient education and rehearsal could become standard practice in radiology, turning what was once an intimidating, mysterious process into one that patients approach with familiarity and confidence. Ultimately, by improving the emotional and procedural readiness of patients, we improve the quality of care itself. We believe this work lays important groundwork for that vision, and we anticipate that immersive preparatory tools will play an increasingly prominent role in making healthcare experiences more patient-friendly and successful.

\bibliographystyle{abbrv-doi-hyperref-narrow}
\bibliography{template}

\begin{thebibliography}{10}
\renewcommand*{\sfdefault}{PTSansNarrow-TLF}

\bibitem{Alanazi2022}
A.~Alanazi, F.~Ashour, H.~Aldosari, and B.~Aldosari.
\newblock The impact of virtual reality in enhancing the quality of life of pediatric oncology patients.
\newblock {\em Stud. Health Technol. Inform.}, 289:477--480, 2022.

\bibitem{Bay2024}
B.~Bay, N.~Voyvoda, and M.~Arifo\u{g}lu.
\newblock An initial investigation into the use of virtual reality (vr) glasses on self-reported pain perception during mammography.
\newblock {\em Radiography}, 30(4):1363--1367, 2024.

\bibitem{Brown2018}
R.~K. Brown, S.~Petty, S.~O'Malley, J.~Stojanovska, M.~S. Davenport, E.~A. Kazerooni, and D.~Fessahazion.
\newblock Virtual reality tool simulates mri experience.
\newblock {\em Tomography}, 4(3):95--98, 2018.

\bibitem{Dewey2007}
M.~Dewey, T.~Schink, and C.~Dewey.
\newblock Claustrophobia during magnetic resonance imaging: cohort study in over 55,000 patients.
\newblock {\em J. Magn. Reson. Imaging}, 26(5):1322--1327, 2007.

\bibitem{Grilo2017}
A.~M. Grilo, L.~Vieira, E.~Carolino, C.~Oliveira, C.~Pacheco, M.~Castro, and J.~Alonso.
\newblock Anxiety in cancer patients during $^{18}$f-fdg pet/ct low dose: A comparison of anxiety levels before and after imaging studies.
\newblock {\em Nurs. Res. Pract.}, 2017:3057495, 2017.

\bibitem{Iosca2025}
N.~Iosca, G.~Busto, Y.~Wandael, A.~Barra, M.~Rossi, I.~Morelli, A.~Pirrera, I.~Desideri, R.~Ricci, L.~Livi, and D.~Giansanti.
\newblock A mixed scoping and narrative review of immersive technologies applied to patients for pain, anxiety, and distress in radiology and radiotherapy.
\newblock {\em Diagnostics}, 15(17):2174, 2025.

\bibitem{Janecky2025}
D.~Janeck{\'{y}}, E.~Ku{\v{c}}era, O.~Haffner, Z.~Ko{\v{s}}utzk{\'{a}}, and P.~Marti{\v{s}}.
\newblock The use of mixed reality for exposure therapy to improve phobia handling.
\newblock {\em Int. J. Cogn. Behav. Ther.}, 18(2):201--240, 2025.

\bibitem{Jimenez2018}
Y.~Jimenez, S.~Cumming, W.~Wang, K.~Stuart, D.~Thwaites, and S.~Lewis.
\newblock Patient education using virtual reality increases knowledge and positive experience for breast cancer patients undergoing radiation therapy.
\newblock {\em Support. Care Cancer}, 26(8):2879--2888, 2018.

\bibitem{Katz1994}
R.~Katz, L.~Wilson, and N.~Frazer.
\newblock Anxiety and its determinants in patients undergoing magnetic resonance imaging.
\newblock {\em J. Behav. Ther. Exp. Psychiatry}, 25(2):131--134, 1994.

\bibitem{Lange2023}
S.~Lange, W.~M{\k{e}}drzycka-D{\k{a}}browska, and A.~Ma{\l}ecka-Dubiela.
\newblock Patient experience during contrast-enhanced computed tomography examination: Anxiety, feelings, and safety.
\newblock {\em Safety}, 9(4):69, 2023.

\bibitem{Lastrucci2024}
A.~Lastrucci, Y.~Wandael, A.~Barra, R.~Ricci, G.~Maccioni, A.~Pirrera, and D.~Giansanti.
\newblock Exploring augmented reality integration in diagnostic imaging: Myth or reality?
\newblock {\em Diagnostics}, 14(13):1333, 2024.

\bibitem{Malviya2000}
S.~Malviya, T.~Voepel-Lewis, O.~Eldevik, D.~Rockwell, J.~Wong, and A.~Tait.
\newblock Sedation and general anaesthesia in children undergoing mri and ct: adverse events and outcomes.
\newblock {\em Br. J. Anaesth.}, 84(6):743--748, 2000.

\bibitem{Melendez1993}
J.~C. Melendez and E.~McCrank.
\newblock Anxiety-related reactions associated with magnetic resonance imaging examinations.
\newblock {\em JAMA}, 270(6):745--747, 1993.

\bibitem{Munn2015}
Z.~Munn, S.~Moola, K.~Lisy, D.~Riitano, and F.~Murphy.
\newblock Claustrophobia in magnetic resonance imaging: a systematic review and meta-analysis.
\newblock {\em Radiography}, 21(2):e59--e63, 2015.

\bibitem{Murphy1997}
K.~J. Murphy and J.~A. Brunberg.
\newblock Adult claustrophobia, anxiety and sedation in mri.
\newblock {\em Magn. Reson. Imaging}, 15(1):51--54, 1997.

\bibitem{Parsons2008}
T.~D. Parsons and A.~A. Rizzo.
\newblock Affective outcomes of virtual reality exposure therapy for anxiety and specific phobias: a meta-analysis.
\newblock {\em J. Behav. Ther. Exp. Psychiatry}, 39(3):250--261, 2008.

\bibitem{Schaake2024}
R.~Schaake, I.~Leopold, A.~Sandberg, B.~Zenk, L.~Shafer, D.~Yu, X.~Lu, S.~Theingi, A.~Udongwo, G.~Cohen, D.~Winkel, D.~Robertson, O.~M. van Delden, and K.~P. van Lienden.
\newblock Virtual reality for the management of pain and anxiety for ir procedures: A prospective, randomized, pilot study on digital sedation.
\newblock {\em J. Vasc. Interv. Radiol.}, 35(8):825--833.e2, 2024.

\bibitem{Sun2020}
Y.~Sun, Y.~Sun, Y.~Qin, Y.~Zhang, H.~Yuan, and Z.~Yang.
\newblock 'virtual experience' as an intervention before a positron emission tomography/ct scan may ease patients' anxiety and improve image quality.
\newblock {\em J. Med. Imaging Radiat. Oncol.}, 64(5):641--648, 2020.

\bibitem{Toussaint2021}
L.~Toussaint, Q.~A. Nguyen, C.~Roettger, K.~Dixon, M.~Offenb{\"{a}}cher, N.~Kohls, J.~Hirsch, and F.~M. Sirois.
\newblock Effectiveness of progressive muscle relaxation, deep breathing, and guided imagery in promoting psychological and physiological states of relaxation.
\newblock {\em Evid. Based Complement. Alternat. Med.}, 2021:5924040, 2021.

\bibitem{Tugwell2018}
J.~R. Tugwell, N.~Goulden, and P.~Mullins.
\newblock Alleviating anxiety in patients prior to mri: a pilot single-centre single-blinded randomised controlled trial to compare video demonstration or telephone conversation with a radiographer versus routine intervention.
\newblock {\em Radiography}, 24(2):122--129, 2018.

\bibitem{yang2024can}
Y.~Yang, E.~A. Corona, B.~L. Daniel, and C.~Leuze.
\newblock Can a novel virtual reality simulator, developed for standalone hmds, effectively prepare patients for an mri examination?
\newblock In {\em 2024 IEEE Conference on Virtual Reality and 3D User Interfaces Abstracts and Workshops (VRW)}, pp. 1037--1038. IEEE, 2024.

\bibitem{yang2025vr}
Y.~Yang, M.~Guo, E.~A. Corona, B.~Daniel, C.~Leuze, and F.~Baik.
\newblock Vr mri training for adolescents: A comparative study of gamified vr, passive vr, 360 video, and traditional educational video.
\newblock {\em arXiv preprint arXiv:2504.09955}, 2025.

\end{thebibliography}
\end{document}